\documentclass{kapedbk}
\usepackage{graphicx,amsmath,amsfonts,edbkps}
\normallatexbib

\begin{document}
\articletitle{Decoherence in Disordered Conductors at Low
Temperatures, the effect of Soft Local Excitations}

\chaptitlerunninghead{Low-Temperature Decoherence in Disordered
Conductors, soft local excitations}

\author{Y. Imry}
\affil{Weizmann Institute of Science, Department of Condensed
Matter Physics, Rehovot 76100, Israel.}
\email{fnimry@weizmann.ac.il}
\author{Z. Ovadyahu}
\affil{The Racah Institute of Physics, Hebrew University, 91904
Jerusalem, Israel.} \email{zvi@vms.huji.ac.il}

\and 

\author{A. Schiller}
\affil{The Racah Institute of Physics, Hebrew University, 91904
Jerusalem, Israel.} \email{avraham@schiller1.phys.huji.ac.il}

\begin{abstract}
The conduction electrons' dephasing rate, $\tau_{\phi}^{-1}$, is
expected to vanish with the temperature. A very intriguing
apparent saturation of this dephasing rate in several systems was
recently reported at very low temperatures. The suggestion that
this represents dephasing by zero-point fluctuations has generated
both theoretical and experimental controversies. We start by
proving that the dephasing rate must vanish at the $T\rightarrow
0$ limit, unless a large ground state degeneracy exists. This
thermodynamic proof includes most systems of relevance and it is
valid for any determination of $\tau_{\phi}$ from {\em linear}
transport measurements. In fact, our experiments demonstrate
unequivocally that indeed when strictly linear transport is used,
the apparent low-temperature saturation of $\tau_{\phi}$ is
eliminated. However, the conditions to be in the linear transport
regime are more strict than hitherto expected. Another novel
result of the experiments is that introducing heavy nonmagnetic
impurities (gold) in our samples produces, even in linear
transport, a shoulder in the dephasing rate at very low
temperatures. We then show theoretically that low-lying local
defects may produce a relatively large dephasing rate at low
temperatures. However, as expected, this rate in fact vanishes
when $T \rightarrow 0$, in agreement with our experimental
observations.
\end{abstract}

\vspace{2cm}

\section{Introduction}

\label{Introduction}

Electronic quantum effects in mesoscopic~\cite{book} and in
disordered conductors~\cite{LR} are controlled by the conduction
electrons' dephasing~\cite{Fey} rate, $\tau _{\phi }$, which is
expected to vanish with the temperature~\cite{AAK,SAI}. A very
intriguing apparent saturation of this dephasing rate in several
systems was recently reported~\cite{webb} at very low
temperatures. Serious precautions~\cite{webb} were taken to
eliminate experimental artifacts. It was speculated that such a
saturation of the dephasing rate when $T\rightarrow 0$ might
follow from interactions with the zero-point motion of the
environment. These speculations have received apparent support
from calculations by Golubev and Zaikin~\cite{zaikin}, which
generated a major controversy~\cite{ale} in the recent literature.
Interestingly, however, this issue had appeared already in 1988,
and good arguments against dephasing by the zero-point motion
have already been given then~\cite{RSS}. Moreover, these results
were in disagreement with other experiments, for example, by
Khavin {\em et al.}~\cite{kha}. More recent experiments~\cite{norm}
showed that in {\it some} cases the presence of trace magnetic
impurities, even on the ppm level, caused the apparent saturation
of the low-temperature dephasing. Similar effects may exist for
models~\cite{IFS,Kozub} with low-lying two-level systems
(TLS)~\cite{Ov84}, where an apparent saturation of the
low-temperature dephasing rate may occur (which will, however,
be eliminated at the $T\rightarrow 0$ limit). For the case of
magnetic impurities, such an elimination of the dephasing rate
will occur if and when Kondo screening of the magnetic moments
or their freezing into a spin-glass state takes place.

In fact, it is physically clear that since dephasing must be
associated with a change of the environment state~\cite{SAI},
it cannot happen as $T\rightarrow 0$, except when a large
ground-state degeneracy occurs. In that limit neither the electron
nor the environment has any energy to exchange. This is a very general
statement; the only physical input needed for it to hold true is that
both the interfering particle and its environment are in equilibrium
at the temperature $T$ (which we then let approach zero). This is so
because the linear transport under consideration is rigorously
determined by equilibrium dynamic correlation functions. Obviously,
nothing prevents a high-energy particle far from equilibrium to
thermalize with the ($T\rightarrow 0$) bath by giving it energy,
losing its phase coherence in the process. Therefore, dephasing
of a particle which is far from equilibrium with a $T=0$ bath is, of
course, possible. As will be discussed later in this article, an
example for such a situation occurs when the energy of the conduction
electrons exceeds the thermal one, due to a voltage bias larger
than $k_{B}T$ and slow relaxation~\cite{Zvi01}.

In the theoretical part of the present paper we do not attempt to
settle the important question of where have the calculations
leading to $T\rightarrow 0$ dephasing without magnetic impurities
or TLS gone wrong~\cite{vD}. We shall start by converting the
above physical argumentation for the lack of $T\rightarrow 0$
dephasing into a more rigorous one. We shall note that, like many
other physical properties, the dephasing rate can be expressed in
terms of correlation functions of the conduction electron and
those of the environment with which it interacts~\cite{book}.
Using very general properties of these correlators~\cite{vH,Uri},
which are almost always valid, we prove following
Refs.~\cite{CI,isqm,IFS} that the dephasing rate has to vanish at
the $T\rightarrow 0$ limit, unless a large ground-state degeneracy
exists. Such a degeneracy may be brought about, e.g., by free
uncompensated magnetic impurities at a vanishing magnetic field.
Because these magnetic moments will typically be screened or
frozen when $T\rightarrow 0$, the proof encompasses most systems
of relevance. Since it employs mainly the basic laws of
thermodynamics, the proof is valid for any case in which
$\tau_{\phi }$ is determined from {\it linear} transport
measurements.

Experiments were performed to examine the real-life validity of
the above statement. Our experiments~\cite{Zvi01} demonstrate
unequivocally that indeed when strictly linear transport is used,
the apparent low-temperature saturation of $\tau_{\phi}$ is
eliminated. Extremely small measurement currents had to be used in
order to be in the linear-transport regime (see also
Ref.~\cite{Berg}). These observations, along with the apparent
lack of heating of the conduction electrons (see Ref.~\cite{webb}
and below), pose new and interesting basic questions.

Another novel result of the experiments is that introducing heavy
nonmagnetic impurities (gold) into our samples produces, even in linear
transport, an anomalously large dephasing rate at very low temperatures,
but {\it not} at the $T\rightarrow 0$ limit. We show that low-lying
local defects, as suggested for example in~\cite{IFS,Kozub}, may
produce a relatively large dephasing rate at low temperatures, which
in fact vanishes when $T\rightarrow 0$.

\section{The vanishing of the dephasing rate as $T
\rightarrow 0$: theory}

\vspace{0.5cm}

\subsection{A useful expression for the dephasing rate}

In this section, we shall derive a very useful expression for the
dephasing rate of a ``particle'' coupled to the ``environment''. The
latter, which will also be referred to as ``the bath'', represents
all the degrees of freedom that the particle is coupled to and are
not directly observed in the interference measurement. As we show
below, the dephasing rate can be expressed as
\begin{equation}
1/\tau _{\phi } = {\frac{1}{\hbar ^{2}(2\pi )^{3}Vol}}
                  \int d{\bf q}
          \int\limits_{-\infty }^{\infty }d\omega
          |V_{{\bf q}}|^{2}S_{p}(-{\bf q},-\omega)
                           S_{s}({\bf q},\omega ) ,
\label{magic}
\end{equation}
where $V_{{\bf q}}$ is the Fourier transform of the interaction
$V(r)$ between the conduction electron and the particles of the
bath, and $S_{p}(-{\bf q},-\omega)$ and $S_{s}({\bf q},\omega)$
are the dynamic structure factors of the conduction electron and
the bath, respectively. These structure factors, which are the
Fourier transforms of the density-density correlation
functions~\cite{vH}, contain the necessary physical information on
both the particle and the bath~\cite{Uri}. A subtle relevant
example is provided by the case where the particle and the bath
are identical fermions. The Pauli principle constraints are
automatically taken into account by using the $S_{p}(-{\bf
q},-\omega)$ of the particle {\it in the presence of the bath}.
These structure factors are in principle known for models of
interest. They can, for example, be obtained from the dissipative
part of the linear response function $1/\epsilon ({\bf q},\omega
)$, by using the fluctuation-dissipation theorem. The physical
meaning of the expression in Eq.~\ref{magic} is simply that the
rate of creating (or annihilating) {\it any} excitation in the
environment is the sum of these rates for all $(q,\omega)$
channels.

The dynamic structure factor is well known for a diffusing electron.
In the classical limit, $\hbar \omega \ll k_{B} T$, it is given by
a Lorentzian of width $Dq^2$. The low-temperature case~\cite{Uri}
will be discussed later on. Replacing $S_{s}(q,\omega)$ with the
dynamic structure factor of the electron gas, given to leading order
by $\hbar q^{2} \omega Vol/2(2 \pi)^{3} e^{2} \sigma$ (see, e.g.,
Eqs.~3.28 and 3.44 of Ref.~\cite{IGB}), allows for an extremely
simple calculation of the dephasing rate by electron-electron
interactions, which reproduces the results of Ref.~\cite{AAK}.

For a derivation of the basic equation~\ref{magic}, we start with
a direct-product state of the particle and the environment,
$|im\rangle \equiv |i\rangle \otimes |m\rangle $, and evaluate
the rate of transitions into {\it all different} possible states,
$|jn\rangle$, using the golden rule. In other words, at any
later time $t$, the state of the total system evolves into
\begin{equation}
\Psi (t) = A \left [ |im\rangle +
                     \sum_{j,n}\alpha _{jn}(t)|jn\rangle \right ],
\end{equation}
where $A$ is a normalization factor. The transition probability from
$|im\rangle$ is simply $|A|^{2}\sum_{j,n}|\alpha _{jn}(t)|^{2}$. At
times larger than microscopic, the transition rate out of the initial
state is well-known to be given by (see, for example, Ref.~\cite{Merz})
\begin{eqnarray}
&& 1/\tau_{out} = \frac{2\pi }{\hbar }\sum_{j,n} |\langle
im|V|jn\rangle|^{2} \int\limits_{-\infty }^{\infty }d(\hbar \omega
) \nonumber\\ \noindent &&\delta (E_{p,j}-E_{p,i}-\hbar \omega )
\delta (E_{s,n}-E_{s,m}+\hbar \omega ). \label{rate}
\end{eqnarray}
Here the last integral represents the (joint) density of final states,
$|jn\rangle $, having the same energy as the initial one, $|im\rangle $.
The matrix elements in the last equation are easily evaluated from the
Fourier representation of the interaction $V(r)$:
\begin{equation}
\langle im|V|jn\rangle = (2\pi )^{-3}\sum_{s}\int d{\bf q}V_{{\bf q}}
               \langle i|e^{i{\bf q}\cdot {\bf r}_{p}}|j\rangle
               \langle m|e^{-i{\bf q}\cdot {\bf r}_{s}}|n\rangle ,
\end{equation}
where the index $s$ runs over the particles in the bath. The absolute
value squared of this matrix element consists of ``diagonal'' terms
(${\bf q}={\bf q}^{\prime }$) which are positive, and ``nondiagonal''
terms (${\bf q}\neq {\bf q}^{\prime }$) whose phases are random. An
important step is now to average the result over, for example, the
impurity ensemble. This will eliminate all the nondiagonal terms,
leaving only the diagonal ones. We now introduce a thermal averaging
over the initial state, $|im\rangle $, by summing over $i$ and
$m$, with the factorized weight of that initial state, $P_{p,i}P_{s,m}$,
in obvious notation. It is immediately recognized that the integral
in Eq.~\ref{rate} contains the product of the dynamic structure
factors of the particle and the environment. As a result, we obtain
Eq.~\ref{magic} for the rate $1/\tau_{out}$.

We emphasize that this result is exact within the golden-rule
formulation, which fully captures the decay of a given initial state
into a continuum. Based on a single-particle picture, it equally
applies to low-energy quasiparticle excitations in a Fermi liquid.
Hence this result extends well beyond perturbation theory for the
{\em bare electrons} in the system. It does not apply to systems
that develop a non-Fermi-liquid ground state.

Using fashionable terminology, $1/\tau_{out}$ is the rate at which the
particle gets ``entangled'' with the environment. In most situations,
$1/\tau_{out}$ is identical to the dephasing rate $1/\tau_{\phi}$.
Important exceptions having to do with the infrared behavior of the
integral in Eq.~\ref{magic}, relevant at lower dimensions, were
discussed in Refs.~\cite{SAI,book}.

\subsection{Proof that the dephasing rate vanishes at the
            $T \rightarrow 0$ limit}

As discussed above, an important advantage of the present
formulation is that all the relevant physical information is
contained within the correct dynamic structure factors of both the
particle and the environment. For example, at $T=0$, when the
electron is diffusing on the Fermi surface, it can not lower its
energy. Hence its dynamic structure factor automatically vanishes
for positive frequencies~\cite{Uri}, as does the dynamic structure
factor of the electron gas at $T = 0$~\cite{IGB}. These facts are
guaranteed by the detailed balance condition:
\begin{equation}
S(-{\bf q},-\omega) = e^{\beta \hbar \omega}S({\bf q},\omega) .
\label{DB}
\end{equation}
More generally, because of the occurrence of $\omega$ and $-\omega$ in
the two dynamic structure factors, the integrand in Eq.~\ref{magic}
vanishes in general at any $\omega$ at the $T \rightarrow 0$ limit,
see Fig.~1. In mathematical terms, it has ``no support''.

\begin{figure}[ht]
\vskip.4in
\includegraphics[width=.8\textwidth]{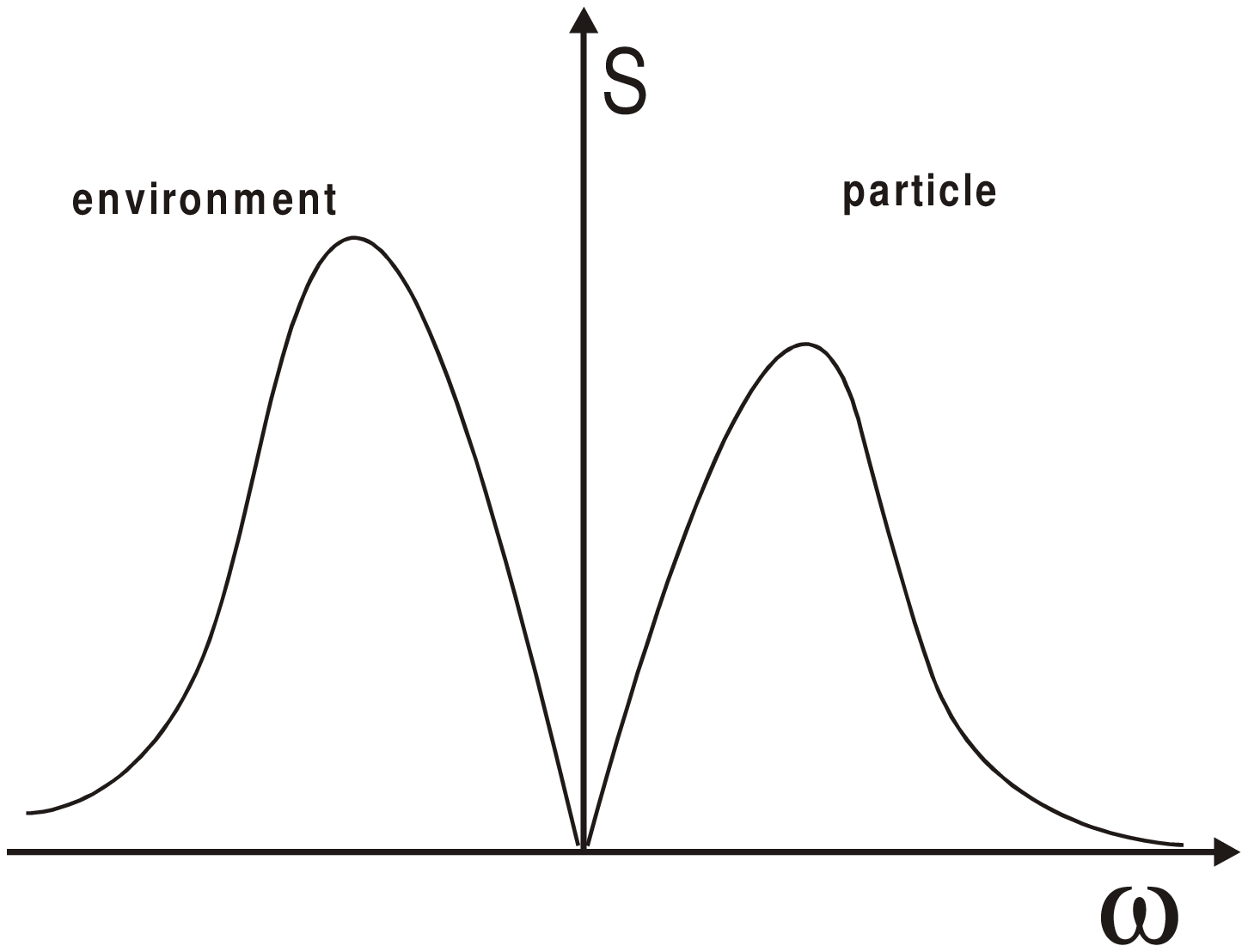}
\caption{The two structure factors appearing in
Eq.~\protect\ref{magic}
         as functions of $\protect\omega$, for $T = 0$ (schematic sketch).
         Note that the product vanishes everywhere.}
         \label{fig1}
\end{figure}

Therefore, except for the unusual case where the environment has a
massive ground-state degeneracy, the dephasing rate must vanish at
zero temperature~\cite{IFS,CI,isqm}. This fact follows directly from
Eq.~\ref{rate} also prior to carrying out the impurity-ensemble
averaging. If both the particle and the environment start at their
lowest states, it is impossible that both $E_{p,j} > E_{p,i}$ and
$E_{s,n} > E_{s,m}$. All this is guaranteed by basic thermodynamics.
The only exception is when a large ground-state degeneracy exists
in the environment, a situation which is very rare indeed, because
such a degeneracy will typically be lifted by some perturbation
that exists in the system.

As mentioned above, this formulation breaks down in the case
of a non-Fermi-liquid ground state, when the elementary excitations
of the system are not of single-particle nature. A notable example
is provided by the two-channel Kondo effect, where
single-particle--to--single-particle
scattering is absent on the Fermi level at $T = 0$.
In fact, the corresponding $S$ matrix has no matrix element to any
outgoing state containing arbitrary finite numbers of particle-hole pair
excitations~\cite{ML97}. This means that any scattering at $T = 0$
necessarily leaves its mark on the environment, resulting in a
finite zero-temperature dephasing rate~\cite{ZvDR}. We emphasize,
however, that the two-channel Kondo effect itself stems from a
ground-state degeneracy that cannot be lifted. Hence a macroscopic
ensemble of two-channel Kondo impurities likewise requires a
macroscopic degeneracy of the ground state.

Thus, the ``standard model'' of linear transport in disordered
metals (in which the defects are strictly frozen) gives, as
expected, an infinite $\tau_{\phi }$ at $T=0$. On the other hand,
there may be other physical ingredients that can make $\tau_{\phi}$
relatively short at very low temperatures (but still {\it divergent}
at the $T\rightarrow 0$ limit), {\it without} contradicting any
basic law of physics. What is needed is an abundance of low-energy
modes in the environment. A simple model for such modes was
suggested in Ref.~\cite{IFS}. Its physics is reminiscent of models
for $1/f$ noise, but the relevant frequencies here are in the
gigahertz range and above. That type of model is a particular one,
and its requirements may or may not be satisfied in real samples.
However, other models with similar dynamics might exist as well.
We reiterate that such models do {\it not} imply dephasing by
zero-point fluctuations. The explanation for the large
low-temperature dephasing rate is certainly not universal. In some
cases this extra low-temperature dephasing depends on sample
preparation, and on extremely small concentrations of stray magnetic
impurities~\cite{norm}. In other cases~\cite{Zvi01} ``nonequilibrium''
behavior, i.e., the sample {\it not} being in the linear-transport
regime, is the relevant issue.

\section{Experimental results}
\label{exp}

Recent experiments on the behavior of the phase coherent time in
indium-oxide films showed some intriguing features. These results
did not confirm the claim~\cite{webb} that there is `inherent
saturation' of $\tau_{\phi}$. In fact, in all cases $\tau_{\phi}$
diverged when $T\rightarrow 0$. On the other hand, several aspects
of the data reproduce the findings of Mohanty {\em et al}. Most
importantly, over a considerable range of measurement conditions
(in particular, the electric field $F$ applied in the
magneto-resistance measurements), the dephasing rate was
$T$-independent below $1$K, while the resistance was
$T$-dependent. The latter suggests that {\it heating} is not a
serious problem in this range of fields, a conclusion reached by
Mohanty {\em et al.} based on the same observation. We also agree
with these authors that external noise is not likely to be the
source of the apparent saturation. In the indium-oxide films,
however, it appears that the saturation is due to {\it
non-equilibrium} effects, namely, when the conductance is no
longer given by the standard second-order current-current
correlation function. Indeed, it was shown that the problem of
apparent saturation disappeared when sufficiently small bias
conditions were employed. It was also shown that in order to be in
the linear-response regime, the electric field used in the
magneto-resistance (MR) measurements must be smaller than $F_{c}=
k_{B}T/eL_{er}$, where $L_{er}$ is the energy relaxation length.
The energy relaxation length $L_{er}$ is the spatial scale over
which the electrons lose their excess energy (gained from their
motion in an electric field) to the environment. This length
should {\it not} be confused with $L_{\phi }$, which is the
phase-coherent diffusion length. Except when dominated by
electron-phonon scattering, $L_{er}$ varies much faster with $T$
than $L_{\phi }$, and may attain macroscopic values at low
temperatures~\cite{Gersh}. In the pure In$_{2}$O$_{3-x}$ samples,
for example, $L_{er}$ reached values of a few mm's below $1$K.

Another intriguing finding in our studies is the behavior of the
dephasing rate versus temperature of the Au-doped samples.
Figure~3 illustrates this behavior for one such sample that was
extensively studied. These measurements were all performed in the
linear-response regime, which was much easier to achieve than in
the pure In$_{2}$O$_{3-x}$ samples due to the relatively short
$L_{er}$. Note that the dephasing rate is well behaved for $T >
2$K, and vanishes as $T\rightarrow 0$ (based on data for $T <
0.6$K). The intermediate temperature regime reveals, however, an
anomaly; $\tau_{\phi }^{-1}$ seems to be almost independent of
temperature. In fact, if the measurements were carried out only
down to $T = 0.6$K, one might have concluded that $\tau_{\phi }$
has saturated! This behavior was observed in all our Au-doped
samples (with doping levels of 1-3\%), and it illustrates a new
type of an apparent saturation problem. The overall shape of
$\tau_{\phi }^{-1}(T)$ is somewhat similar to the respective
behavior observed in Au films doped with Fe \cite{Berg}, and in Cu
films doped with Cr~\cite{Hasen}. Both are well-known Kondo
systems, and the ''hump\textquotedblright\ observed in their
$\tau_{\phi }^{-1}(T)$ data was indeed interpreted as extra
dephasing due to the Kondo effect. When we tried to repeat the
analysis of these authors on our data, we encountered a number of
difficulties. In the first place, to fit the excess dephasing rate
with the formulae used by these authors required a spin of the
order of 10 (rather than $\frac{1}{2}$ in their case), which makes
no physical sense. More importantly, we failed to detect any
independent evidence for magnetic impurities (above 1ppm), either
in the sample or in the Au material that was used for
doping~\cite{Zvi01}. In addition, there is strong evidence against
dephasing by magnetic impurities in the MR data themselves.
Consider the MR data shown in Fig.~2. The values of
$\tau_{\phi}^{-1}(T)$ are obtained by fitting MR data to
weak-localization theory, which usually is based on data taken at
small magnetic fields. In the graphs of Fig.~2, however, we
deliberately extended the MR measurements to include much larger
fields. Note that in both graphs data are shown up to fields that
are high enough to cause significant spin polarization (the Zeeman
energy exceeds $k_{B}T$). Yet, a nearly perfect fit to the theory
(dashed black lines) is obtained using one value of $\tau_{\phi}$
for each temperature. If there were a contribution from a
spin-flip mechanism (as one may expect from the presence of
magnetic impurities), it would be impossible to fit the low-field
data (namely, for $H < k_{B}T/\mu_{B}g$) with the same
$\tau_{\phi}$ as the one necessary for $H > k_{B}T/\mu_{B}g$. The
difference that might be expected is illustrated in the top graph
of Fig.~2 by the dotted line. The latter represents the MR that
ought to be observed when the extra contribution to dephasing by
the alleged Kondo impurities is suppressed by $H$. It would
therefore appear that the anomaly represented by the
``hump\textquotedblright\ around $T = 0.6$K in Fig.~3 is not due
to the usual spin-flip scattering, resulting from the presence of
magnetic moments.

\begin{figure}[ht]
\vskip.5in
\includegraphics[width=.8\textwidth]{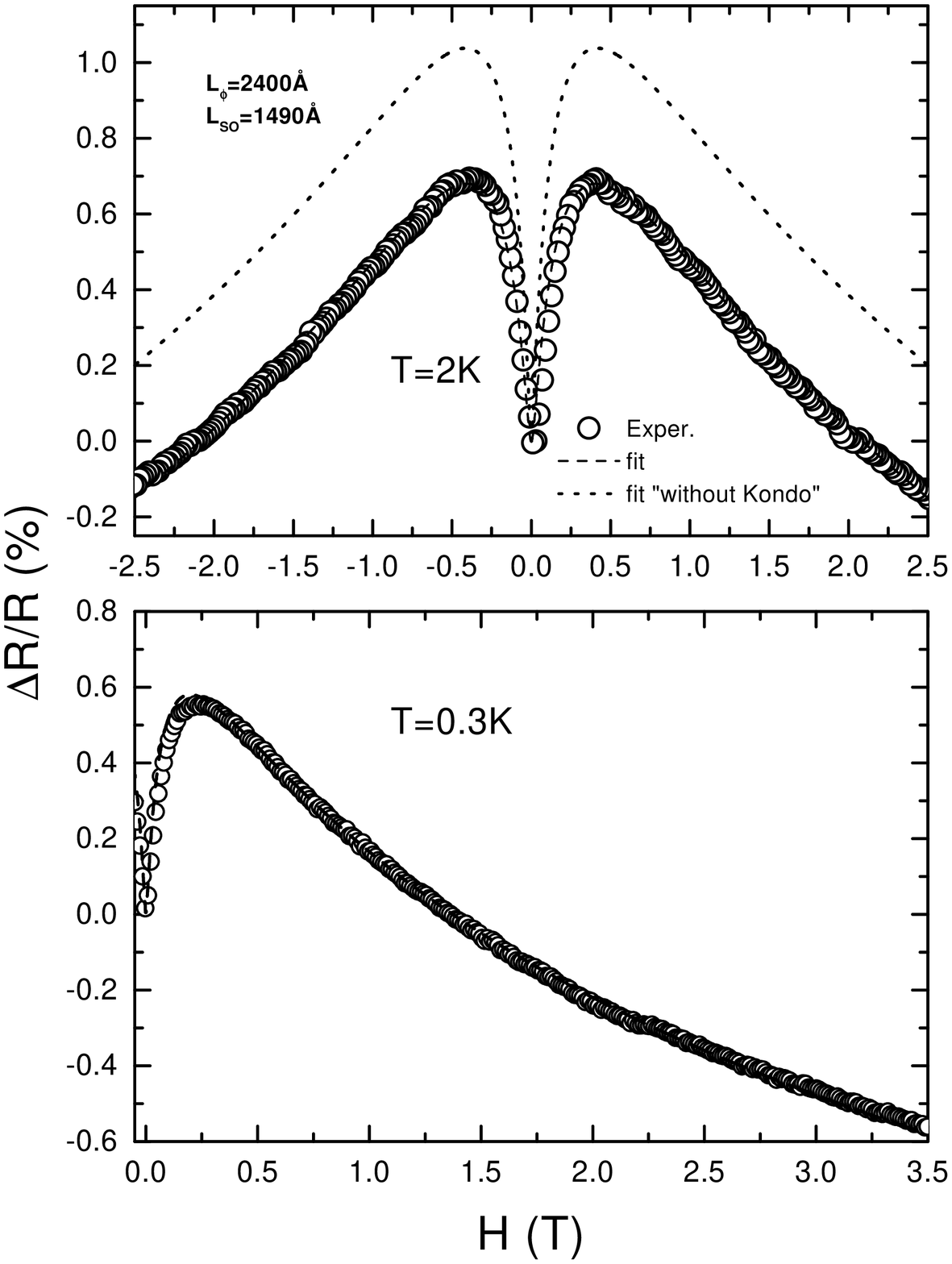}
\caption{MR for In$_{2}$ O$_{3-x}$:Au sample (thickness 200$\AA$
with 2$\%$ Au). Dashed lines are fits to theory using a {\em
single} $\tau_{\phi }$ for each of the temperatures shown (one
above and one below the anomaly).} \label{fig2}
\end{figure}

We shall now attempt to explain the restricted saturation below
$2$K in these samples based on the observation that this anomaly
originates form the inclusion of gold atoms in the In$_{2}$O$_{3-x}$
matrix, and is absent in the pure material. As noted
elsewhere~\cite{Shapir,Kowal,Zvi01}, the Au atoms probably
reside in the oxygen vacancy (or di-vacancy) sites of the
In$_{2}$O$_{3-x}$ (which is typically 10\% oxygen deficient, see
Ref.~\cite{OOK}). Given the chemical inertness of gold, it is not
implausible that a sizeable portion of the Au atoms are loosely
trapped in oxygen di-vacancies, thus acting as local scatterers
with a low characteristic frequency. For simplicity, we model such
a defect as a local TLS having a typical energy $\Delta$ (associated
with two nearly equivalent positions of the Au in the di-vacancy).
The dephasing rate versus temperature due to this model will be
calculated in the next section. It is shown to be consistent with
our experiment in Fig.~3.

\begin{figure}[ht]
\vskip.5in
\includegraphics[width=.8\textwidth]{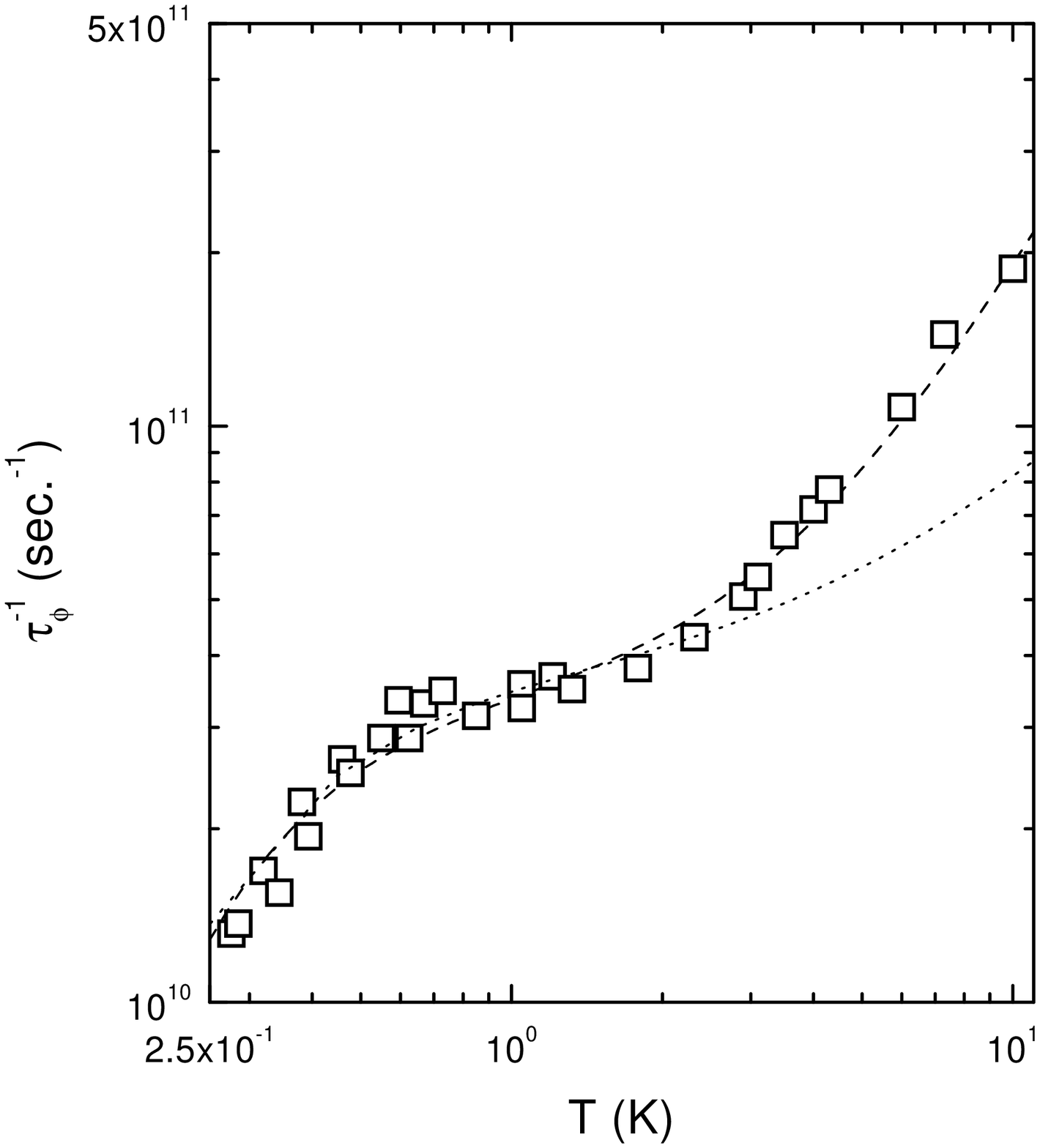}
\caption{Dephasing rate versus temperature for the same
         In$_{2}$O$_{3-x}$:Au sample as in Fig. 2.
         The dotted line is a fit to Eq.~13 with a symmetric well,using
         $n_{s}v_{F}\sigma_{0}=3.4 \cdot 10^{10}$sec$^{-1}$,
         $\Delta =0.3$K, and adding the standard 2d result \cite{AAK}
         for the present situation: $\tau_{\phi}^{-1}
         = 5\cdot 10^{9}\cdot T$ $\cdot $sec$^{-1}$, where T is in degrees K. A better fit
         to the data (squares) may be obtained by using the
         result that would apply were the  film behaving as in 3D \cite{AAK}:
         $\tau_{\phi }^{-1} = 4\cdot 10^{9}\cdot T^{\frac{3}{2}}
         \cdot$ sec$^{-1}$ (where T is again in degrees K),
         for the high-temperature regime (dashed line).}
         \label{fig3}
\end{figure}

\section{A tunnelling model for loosely bound heavy impurities}

\label{models}

In this section, we consider the inelastic scattering of the conduction
electrons from loosely bound defects. The defects are taken for simplicity
to be independent Born-approximation s-wave scatterers, having a scattering
length $a$ and a total scattering cross section $4 \pi a^{2}$. The
differential cross-section for inelastic scattering of a particle with
momentum ${\bf k}$ into an element of solid angle $\Omega$ around the
final momentum ${\bf k^{\prime}}$ is given by~\cite{vH}
\begin{equation}
\frac{\partial^2 \sigma}{\partial \Omega \partial \omega} =
     a^2 S({\bf q},\omega) =
     a^2 \sum_{i,f} P_i |\langle f |e ^{i{\bf q\cdot x}}
     |i\rangle|^2 \delta(\omega - \omega_{if}) ,
\label{x-s}
\end{equation}
where $\hbar \omega$ is the energy transfer, $\hbar {\bf q}$ with
${\bf q = k - k^{\prime}}$ is the momentum transfer, and
$S({\bf q},\omega)$ is the dynamic structure factor of the scatterer.
Here $|i\rangle$ and $|f\rangle$ are the initial and final states of
the scatterer (the former having a probability of $P_{p, i}$), and
$\hbar \omega_{if}$ is their energy difference.

In the tunnelling model we take the scatterer to reside in a
double-minimum potential. The minima are separated by a vector
${\bf b}$, the tunnelling matrix element between the two minima is
$\Omega_0$, and their energy separation is $2B$. By diagonalizing
the $2 \times 2$ problem, one easily finds \cite{67,IFS} that the
separation $2\Delta$ between the ground state and excited state in
the well, $|+\rangle $ and $|-\rangle$, respectively, is given by
\begin{equation}
2\Delta = 2 \sqrt{\Omega_0^2 + B^2} .
\label{sep}
\end{equation}
The above labelling of the states reflects their spatial symmetry
for $B = 0$. The transition matrix element is given in turn by
\begin{equation}
\langle + |e ^{i{\bf q\cdot x}}|i\rangle =
      2i\alpha \beta \sin({\bf q\cdot b}/2)
      \cong i\alpha \beta ({\bf q\cdot b}) ,
\label{ME}
\end{equation}
where $\alpha$ and $\beta$ are the normalized weights in the two
wells. Their product is $\alpha \beta = \Omega_0 / (2 \Delta)$. The
combination $2|\alpha \beta|$ is a symmetry parameter, ranging from
unity for a symmetric well to zero for a very asymmetric one, rendering
the latter ineffective for the inelastic scattering. To get the
second equality in Eq.~\ref{ME}, we used the dipole approximation
${\bf q\cdot b}/2 \ll 1$, which is appropriate for $k_F b \ll 1$.
For simplicity, we took $b$ to be sufficiently large as compared
to the characteristic length of each well. We shall also assume
$\epsilon_F \gg \Delta, k_B T$.

The inelastic cross section for scattering between the two levels
of the tunnelling center is given by
\begin{equation}
\sigma_{in}({\bf q},\omega) = 4\alpha^2 \beta^2a^2
       \sum_{\gamma = \pm}P_{\gamma} \sin^2({\bf q\cdot b}/2)
       \delta (\omega + 2 \gamma \Delta).
\label{scatt}
\end{equation}
Here $P_{\pm}$ are the thermal populations of the $|\pm\rangle$ states:
\begin{equation}
P_{\pm} = \frac {e^{\pm \Delta/(k_BT)}} {2 \cosh(\Delta/(k_BT))}.
\label{pop}
\end{equation}
For simplicity we consider an electron with an initial momentum
${\bf k}$ very close to the Fermi sphere, i.e., $\epsilon_k \ll
k_BT$, where the kinetic energy $\epsilon_k$ is measured relative
to the chemical potential. The total inelastic cross sections for
an upwards/downwards excitation of the TLS are given by

\begin{equation}
\sigma_{\pm} = \alpha^2 \beta^2a^2 P_{\pm}
       \int d\Omega_{{\bf k^{\prime}}} \int d\epsilon_{k^{\prime}}
       ({\bf (k - k^{\prime})\cdot b})^2
       [1 - f(\epsilon_{k^{\prime}})]
       \delta (\epsilon_k - \epsilon_{k^{\prime}} \mp 2 \Delta),
\label{sigma_pm}
\end{equation}
where $d\Omega_{{\bf k^{\prime}}}$ is an element of solid angle
around the final wave vector ${\bf k^{\prime}}$, and $\epsilon_k =
\hbar^2 k^2 /2m$. For clarity we take the initial wave vector
${\bf k}$ to be parallel to ${\bf b}$ for the time being.
Averaging over the direction of ${\bf k}$ will introduce a
numerical factor $\lambda$, which we shall reinstate later on.

To proceed with Eq.~\ref{sigma_pm}, we note that ${\bf (k -
k^{\prime})\cdot b}$ equals $2k_F b \sin^2(\theta /2)$, where
$\theta$ is the angle between ${\bf k^{\prime}}$ and ${\bf b}$.
Performing the angular integration and the integral over the
energy, we obtain
\begin{eqnarray}
&& \sigma_{in, tot} = 16\sigma_{0} (\alpha \beta)^2
       \left [
               P_+ (1- f(\epsilon_k - 2 \Delta)) +
               P_- (1- f(\epsilon_k + 2 \Delta))
       \right] =
\nonumber\\ \noindent &&\frac{4\sigma_{0}(\alpha
\beta)^2}{\cosh^{2}(\Delta/(k_BT))}. \label{sigma_in_tot}
\end{eqnarray}
The prefactor $\sigma_{0}$ in Eq.~\ref{sigma_in_tot} is given by
$\sigma_{0} \equiv \frac{\pi}{3} \lambda a^2 (k_F b)^2$, and is
expected to be of the order of the square of a small fraction of
an Angstrom. For a concentration $n_s$ of the soft impurities, the
rate for inelastic scattering is thus given by
\begin{equation}
\frac{1}{\tau_{in,s}} = \frac{4(\alpha \beta)^2 n_s v_F
        \sigma_0}{\cosh^{2}(\Delta/(k_BT))},
\label{result}
\end{equation}
where $4(\alpha \beta)^2 = 1$ in the symmetric case ($B = 0$).
Note that the situation here is rather distinct from the one for the
electron-electron scattering with disorder, where the scattering is
dominated by small $q^{\prime}s$ (the infrared regime). Since the
scatterers are short ranged, the important range of $q$ is
$q \ell \gg 1$ for $k_F \ell \gg 1$, as in Ref.~\cite{IFS}. In this
range, the dynamics of the electrons is effectively ballistic. For
the same reason, the inelastic rate and dephasing rate are essentially
equal~\cite{Blanter}.

The parameters of the various TLS's within the system, are often
distributed. Reasonable distributions are~\cite{IFS}: a uniform
distribution for $B$ in the range $0 \leq B \leq B_{max}$, and a
$1/\Omega_0$ distribution for $\Omega_0$, between $\Omega_{min}$
and $\Omega_{max}$. The latter distribution follows by taking
$\Omega_0$ to be the exponential of a large negative, uniformly
distributed quantity in the corresponding range. One generally
expects $\Omega_{max} \ll B_{max}$. Thus, the combined
distribution function reads

\begin{equation} P(B, \Omega_0) = \frac{1}{\Omega_0 B_{max}
                 \ln(\Omega_{max} / \Omega_{min})} .
\label{dist} \end{equation}

\begin{figure}[ht]
\vskip.4in
\includegraphics[width=.8\textwidth]{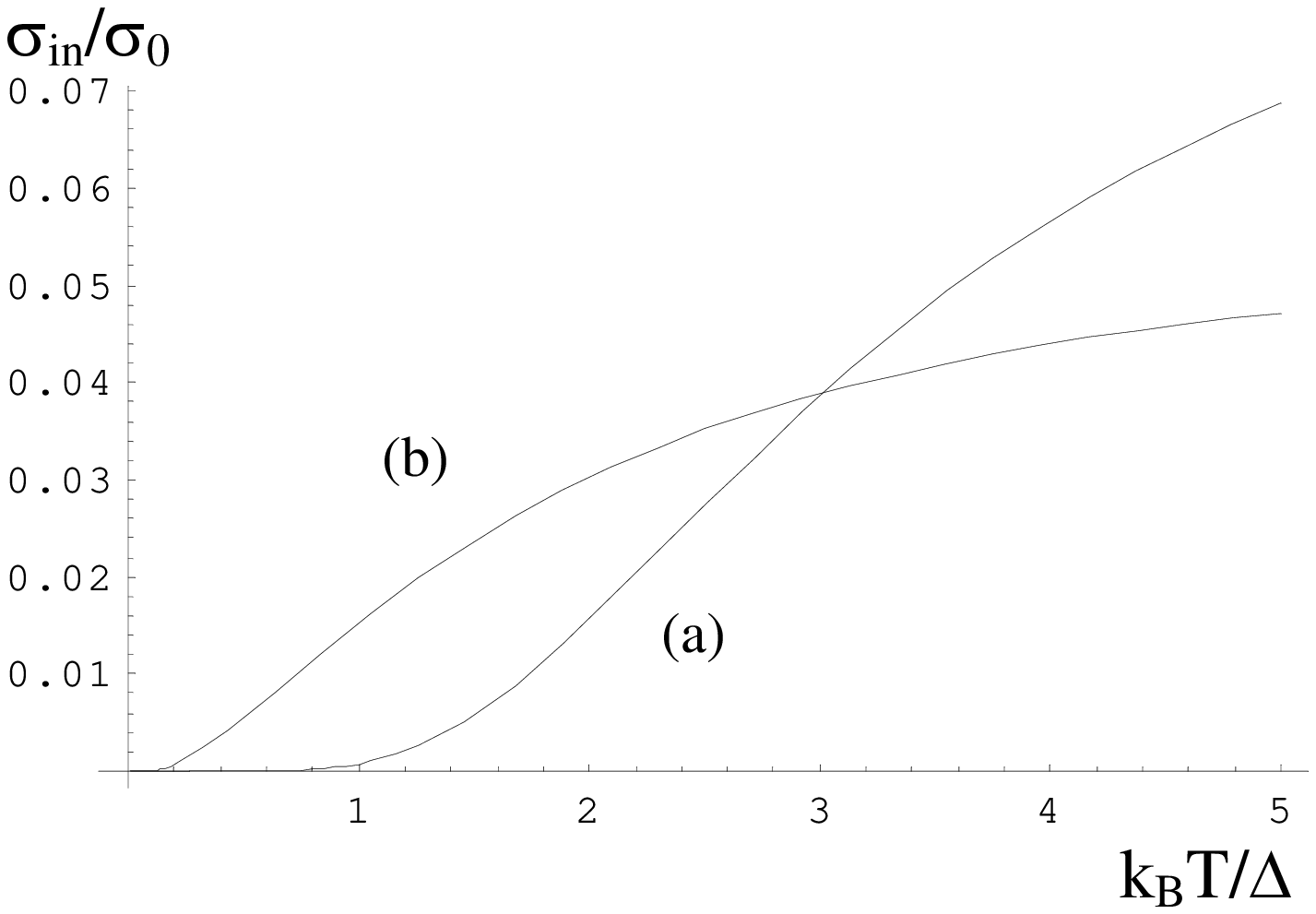}
\caption{The inelastic cross-section of the TLS as a function of
         $k_BT$. (a) A single TLS (Eq.~\ref{sigma_in_tot}) with $B = 3$
         and $\Omega_0 = 1$ (all energies in the same units). (b) The
         cross section averaged over the distribution of Eq.~\ref{dist},
         with $B_{max} = 20$, $\Omega_{min} = 0.2$, and $\Omega_{max} = 2$.
         Note the qualitative similarity between these results and
         the hump of Fig.~3. Adding the electron-electron contribution
         as in Ref.~\protect\cite{AAK} produces a reasonable fit of
         the experimental results with a TLS model,see Fig.~3.}
         \label{fig4}
\end{figure}

The inelastic cross section of a single TLS, and the one averaged
over the distribution of Eq.~\ref{dist}, are depicted in Fig.~\ref{fig4}.
For the TLS distribution of Eq.~\ref{dist}, the following qualitative
behavior of the averaged cross section $\langle \sigma_{in} \rangle$
is found:

\begin{eqnarray}
&& \langle \sigma_{in}\rangle \propto e^{-2\Omega_{min}/(k_{B}T)}
          \;\;\;\;\;\;  for \;\;\;\;\; k_{B}T \ll \Omega_{min};
\nonumber \\
&& \langle\sigma_{in}\rangle \propto T \;\;\;\;\;\;\;\;\;  for
          \;\;\;\;\; \Omega_{min} \ll k_{B} T \ll \Omega_{max};
\nonumber \\
&&\langle\sigma_{in}\rangle \propto const. \;\;\;\;\;\;\;\;\;\;\;\;\;\;\;\;
          for \;\;\;\;\; k_{B}T \gg \Omega_{max}.
\label{resultsav}
\end{eqnarray}

\noindent The behaviors of Eq. \ref{resultsav} are in agreement
with curve (b) of Fig.~\ref{fig4}.

Strictly speaking, these results hold only for temperatures
sufficiently low so that the higher levels of the double-minimum
well are thermally inaccessible. The constant nature of the
inelastic rate for $k_{B}T \gg \Delta_{min}$ was invoked in
Ref.~\cite{IFS} to explain the apparent saturation of the
dephasing rate. This necessitates $\Delta_{max} \approx 0.1K$ -
$0.5K$, as would seem appropriate for heavy defects. As pointed
out in Ref.~\cite{IFS}, the dephasing rate will then vanish
linearly with $T$ at lower temperatures. If the lower cutoff
$\Omega_{min}$ exists and is attainable, the TLS dephasing rate
should eventually vanish exponentially, as specified above. To
remove any doubt, we reemphasize that models without a large
enough ground-state degeneracy {\em do not have a saturation of
$\tau_{\phi}$ at the $T \rightarrow 0$ limit.}

\section{Conclusions}
\label{Conc}

We showed that in normal systems, that do not have large ground
state degeneracies, the quasiparticle dephasing rate {\em must}
vanish at the $T \rightarrow 0$ limit.  Abundance of low-energy
excitations can, however, produce a relatively large dephasing rate
at low nonzero temperatures. An appropriate TLS model can explain
an intriguing feature of our experimental results below $1$K,
obtained by controlled addition of heavy impurities. An apparent
low-temperature saturation of the dephasing rate can also be due
to magnetic impurities, as long as their magnetic moments are
uncompensated and unfrozen. We also find experimentally that the
condition to be in the linear-transport regime at very low
temperatures is much more strict then ordinarily expected. Not
reaching the linear-transport regime might also produce an apparent
``nonequilibrium'' saturation. More theoretical work is necessary in
order to fully understand this last result.

\vspace{.3cm}

{\bf Acknowledgements}\newline This research was supported by the
Centers of Excellence Program of the Israel Science Foundation.
Research at the Weizmann Institute was supported in part by a
grant from the German-Israel Foundation (GIF), Jerusalem and by
the Maurice and Gabriella Goldschleger Center for NanoPhysics. We
thank Y. Aharonov, H. Fukuyama, D. Cohen, P. Schwab and A. Stern
for collaborations on related problems. B. Altshuler, I. Aleiner,
N. Argaman, M. Berry, N. Birge,  Y. Levinson, H. Pothier, T.D.
Schultz, A. Stern, R.A. Webb, H.A. Weidenm\"uller, P. W\"olfle and
A. Zaikin are thanked for discussions.

\end{document}